# Photolysis of *n*-butyl nitrite and isoamyl nitrite at 355 nm: A time-resolved Fourier transform infrared emission spectroscopy and *ab initio* study

**Min Ji, Junfeng Zhen, Qun Zhang,[a] and Yang Chen[a]**

*Hefei National Laboratory for Physical Sciences at the Microscale and Department of Chemical Physics, University of Science and Technology of China, Hefei, Anhui 230026, People's Republic of China*

[a] Authors to whom correspondence should be addressed. Electronic addresses: qunzh@ustc.edu.cn and yangchen@ustc.edu.cn

## ABSTRACT

We report on the photodissociation dynamics study of *n*-butyl nitrite and isoamyl nitrite by means of time-resolved Fourier transform infrared (TR-FTIR) emission spectroscopy. The obtained TR-FTIR emission spectra of the nascent NO fragments produced in the 355-nm laser photolysis of the two alkyl nitrite species showed an almost identical rotational temperature and vibrational distributions of NO. In addition, a close resemblance between the two species was also found in the measured temporal profiles of the IR emission of NO and the recorded UV absorption spectra. The experimental results are consistent with our *ab initio* calculations using the time-dependent density functional theory at the B3LYP/6-311G(d,p) level, which indicate that the substitution of one of the two γ-H atoms in *n*-$C_4H_9ONO$ with a methyl group to form $(CH_3)_2C_3H_5ONO$ has only a minor effect on the photodissociation dynamics of the two molecules.

## I. INTRODUCTION

Alkyl nitrites (generally denoted as RONO, where R represents an alkyl moiety) are a class of molecules that are believed to play important roles in photochemistry.[1] Previous studies showed that their dissociation is fast and their product quantum-state distributions carry memory of the initial excitation. Besides, the UV absorption spectra of RONO species with different R−O groups have been found to be quite similar. The first weak absorption band in the wavelength range of 300 – 400 nm was assigned to the $S_1(n\pi^*) \leftarrow S_0$ transition (localized in the −N=O chromophore) that corresponds to the excitation of one of the lone-pair electrons of oxygen atom to the antibonding $\pi^*$ orbital.[2-6] This band exhibits spectral features attributable to the N=O stretching mode in the $S_1$ excited state with different vibrational quantum numbers,[7] and its relevant photodissociation follows the processes: R−O−N=O $(S_0)$ + $h\nu$ → R−O−N=O $(S_1, n^*)$ → R−O + N=O $(X^2\Pi, n, j)$, where $n^*$ denotes the vibrational quantum number for the N=O stretching mode in the $S_1$ complex and $(n, j)$ are the vibrational and rotational quantum numbers of the ground-state NO fragment. The second absorption band, which appears stronger than the first band, originates from the $S_2(\pi\pi^*) \leftarrow S_0$ transition centered around 220 nm.[2] The primary dissociation pathway associated with the two UV absorption bands was found to be due to the cleavage of the weak RO−NO bond (~170 kJ/mol).[8]

While extensive studies have been performed on nitrous acid (HONO),[3-5,9-11] methyl nitrite ($CH_3ONO$),[6,7,12-15] and *tert*-butyl nitrite (($CH_3$)$_3CONO$),[7,16-19] there remains quite meager in the photodissociation studies on *n*-butyl nitrite (*n*-$C_4H_9ONO$) and isoamyl nitrite (($CH_3$)$_2C_3H_5ONO$). To the best of our knowledge, only two recent works on the photodissociation dynamics of $n$-C$_4$H$_9$ONO were reported by Yue $et\ al.$,[2,20] who measured the nascent NO[2] and OH[20] fragments from the photodissociation of gaseous $n$-C$_4$H$_9$ONO using the one-photon laser-induced fluorescence technique.  Nevertheless, no investigations on the photodissociation dynamics of (CH$_3$)$_2$C$_3$H$_5$ONO have hitherto been reported.

In this work, we present a photodissociation dynamics study on both $n$-butyl nitrite and isoamyl nitrite using the time-resolved Fourier transform infrared (TR-FTIR) emission spectroscopic technique.[21,22]  We obtained the TR-FTIR emission spectra of the nascent NO fragments produced in the 355-nm laser photolysis of the two alkyl nitrite species, from which the rotational temperatures and the vibrational distributions of NO for both species were derived from a rovibrational simulation and found to be almost identical.  Such a finding was further supported by the measured temporal profiles of the IR emission of NO, the recorded UV absorption spectra, and high-level $ab\ initio$ calculations.

## II. EXPERIMENTS

Our experiments were carried out under bulk conditions.  The photolysis of $n$-C$_4$H$_9$ONO and (CH$_3$)$_2$C$_3$H$_5$ONO was initiated in a stainless steel flowing cell, and the resultant IR emission from the nascent NO fragments was collected by a pair of gold-coated Welsh mirrors[23] inside the cell and passed through a CaF$_2$ window and a set of collimating lens into a TR-FTIR spectrometer (IFS 66/S, Bruker) that was operated in a step-scan mode.[24,25]

The light source used was the third-harmonics output (355 nm, ~10 ns pulse duration) of a Nd:YAG laser (GCR-170, Spectra-Physics) operating at a 10-Hz repetition rate.  The fluence of the 355-nm laser beam at the detection center of the flowing cell was estimated to be ~70 mJ cm$^{-2}$. Coupled with the TR-FTIR spectrometer was a mercury cadmium telluride detector (MCT, KV 100-1-D-7/190, Kolmar Technologies) with a risetime of ~3.5 ns, and its transient signal was preamplified with a gain factor of ~2 × 10$^4$ V A$^{-1}$ (bandwidth ~50 MHz), followed by further amplification by a factor of 20 (bandwidth ~300 kHz) before being digitized with an external data-acquisition board (PAD82a, 100/200 MHz, Spectrum) at 50-ns intervals. Data were typically averaged over 30 laser pulses at each scan step; 2961 scan steps were performed to yield an interferogram resulting in a spectrum with a resolution of ~1.0 cm$^{-1}$ in the energy region of 1317 – 2633 cm$^{-1}$. To improve the signal-to-noise ratio of the spectrum, 20 consecutive time-resolved spectra were subsequently summed to yield spectra representing emission at 1.0-$\mu$s intervals.

*n*-butyl nitrite (95%, Acros Organics) and isoamyl nitrite (97%, Acros Organics) used in the experiments were degassed through three freeze-pump-thaw cycles in liquid nitrogen and used without further purification. The partial pressures of *n*-butyl nitrite and isoamyl nitrite in the flowing cell at 298 K were maintained at ~20 and ~10 Pa. With Ar purging near the entrance and exit of the photolysis port, the total pressure of the cell was maintained at ~500 Pa.

**III. RESULTS AND DISCUSSION**

The upper trace of Figure 1 shows the partial TR-FTIR emission spectrum (1700 – 2000 cm$^{-1}$, resolution ~1.0 cm$^{-1}$) of the NO fragment recorded in the time interval of 1.0 – 2.0 $\mu$s after laser photolysis of a flowing mixture of *n*-C$_4$H$_9$ONO (~ 20 Pa) and Ar (~ 480 Pa).

Since the nascent NO fragments were produced in their electronic ground state ($X^2\Pi$) under our experimental conditions,[26] we assign this spectrum to the $\upsilon = 1 \rightarrow \upsilon = 0$, $\upsilon = 2 \rightarrow \upsilon = 1$, and $\upsilon = 3 \rightarrow \upsilon = 2$ transitions in the ground $X^2\Pi$ state of NO. Because of a severe spectral congestion among the observed three vibrational bands, we herein label only the band origins, as indicted by ticks in Fig. 1. It is noteworthy that a survey spectrum recorded in the energy range of 0 – 5266 cm$^{-1}$ did not show any other IR emission features that are associated with $n$-$C_4H_9O$, the cofragment of NO.

A simulated spectrum is shown in the lower part of Fig. 1. A Boltzmann function for the rotational distribution of NO was used in the simulation, yielding a rotational temperature of ~330 K and a relative vibrational population of ($\upsilon = 1$) : ($\upsilon = 2$) : ($\upsilon = 3$) = (0.59) : (0.27) : (0.14), which indicates that the nascent NO photofragments were formed to be rotationally "cold" while vibrationally "hot". The simulated spectrum matches nicely with the observed one, which supports our vibrational assignments given in Fig. 1.

It should be noted, however, that each rotational level of the ground vibronic state of NO is split by spin-orbit interaction into two components $F_1(X^2\Pi_{1/2})$ and $F_2(X^2\Pi_{3/2})$ ($\Delta E(F_1 - F_2) \approx 120$ cm$^{-1}$).[27,28] Since the difference of the vibrational frequency ($\Delta\omega_e$) as well as that of the anharmonicity constant ($\Delta\omega_e x_e$) between the $X^2\Pi_{1/2}$ and $X^2\Pi_{3/2}$ states are quite small (~0.16 cm$^{-1}$ and ~0.025 cm$^{-1}$,[29] respectively) compared to our spectral resolution (~1.0 cm$^{-1}$), while what we detected is all from the $\Delta\upsilon = 1$ transitions in the two split ground states ($X^2\Pi_{1/2}$ and $X^2\Pi_{3/2}$), the two spin-orbit components could not be resolved in our observed spectra.

A similar experiment was performed for $(CH_3)_2C_3H_5ONO$. The lower part of Figure 2

shows the partial TR-FTIR emission spectrum (1700 – 2000 cm$^{-1}$, resolution ~1.0 cm$^{-1}$) of the NO fragment recorded in the time interval of 1.0 – 2.0 μs after laser photolysis of a flowing mixture of $(CH_3)_2C_3H_5ONO$ (~ 10 Pa) and Ar (~ 490 Pa).  Since the partial pressure of $(CH_3)_2C_3H_5ONO$ was lowered by a factor of 2 compared to that of *n*-$C_4H_9ONO$ described above, this spectrum is rescaled (2×) for convenience of comparison with the upper part of Fig. 2 that is taken from Fig. 1.  It can be easily seen that the two spectral traces in Fig. 2 resemble one another almost exactly.  Therefore, a rotational temperature of ~330 K and a relative vibrational population of ($v$ = 1) : ($v$ = 2) : ($v$ = 3 ) = (0.59) : (0.27) : (0.14) of the nascent NO fragment, which are virtually identical to those for *n*-$C_4H_9ONO$, can be expected for $(CH_3)_2C_3H_5ONO$.

In addition, we measured the temporal profiles (at a resolution of 200 ns) of the IR emission of the NO fragments coming from both *n*-$C_4H_9ONO$ and $(CH_3)_2C_3H_5ONO$, as shown in Figure 3.  Both traces reaching a maximum at ~0.75 μs were obtained without the 300-kHz-bandwidth amplifier (in order to remove the instrumental influence on the leading edge of the temporal profile), from which no discernable difference could be seen.

We further recorded the electronic absorption spectra in the UV region of 300 – 450 nm for both species in the liquid phase (~1% in ethanol), as shown in Figure. 4.  Again, it is difficult to distinguish the two species from the spectra shown in Fig. 4.  Remarkably, however, it was found that another alkyl nitrite species, *tert*-butyl nitrite $((CH_3)_3CONO)$, exhibits an electronic absorption spectrum with similar spectral features but a rather large wavelength red-shift (~25 nm)[7,30] with respect to that of *n*-butyl nitrite (*n*-$C_4H_9ONO$).

To better understand the above phenomena pertinent to *n*-$C_4H_9ONO$ and $(CH_3)_2C_3H_5ONO$, we carried out *ab initio* calculations using the time-dependent density functional theory (TD-DFT) with GAUSSIAN 03 program.[31] The predicted minimum-energy geometries for $n\text{-}C_4H_9ONO$ and $(CH_3)_2C_3H_5ONO$ in the $S_0$ state are displayed in Figure 5. A selection of relevant geometric parameters calculated at the B3LYP/6-311G(d,p) level for both molecules are listed in Tables 1 and 2, respectively.

In accord with those reported in a very recent work,[32] our calculated results for $n\text{-}C_4H_9ONO$ also showed that all of the C, N, and O atoms are in one plane (denoted as *X-Y* plane). While one of the two γ-H atoms in $n\text{-}C_4H_9ONO$ is substituted with a methyl group to form $(CH_3)_2C_3H_5ONO$, our calculations showed that such a substitution has only a minor effect on the geometry of the main chain of $C_1\text{-}C_2\text{-}C_3\text{-}C_4\text{-}O_5\text{-}N_6\text{-}O_7$, i.e., $C_4$ was found to be still in the *X-Y* plane with $C_3$ and $C_2$ sticking slightly out of plane and $C_1$ slightly into plane: the dihedral angles for $O_7\text{-}N_6\text{-}O_5\text{-}C_4$, $N_6\text{-}O_5\text{-}C_4\text{-}C_3$, $O_5\text{-}C_4\text{-}C_3\text{-}C_2$, and $O_7\text{-}N_6\text{-}O_5\text{-}C_1$ were found to be 0.0º, 179.7º, 175.8º, and -7.9º, respectively. Such a minor effect induced by the substitution of the methyl group can also be seen in Figure 6, which shows almost identical profiles of the highest occupied orbitals (HOMO) as well as the lowest unoccupied orbitals (LUMO) relating to the $S_1 \leftarrow S_0$ transition for $n\text{-}C_4H_9ONO$ and $(CH_3)_2C_3H_5ONO$, i.e., the electronic density for the two HOMO's is localized on the main chain of $C_1\text{-}C_2\text{-}C_3\text{-}C_4\text{-}O_5\text{-}N_6\text{-}O_7$ while that for the two LUMO's primarily on the C−O−N=O core with a planar geometry.

Table 3 lists the vertical excitation energies (VEE) and oscillator strengths (OS) of the $S_1 \leftarrow S_0$ transition for both $n\text{-}C_4H_9ONO$ and $(CH_3)_2C_3H_5ONO$ calculated by TD-DFT at the B3LYP/6-311G(d,p) level based on the optimized geometries shown in Fig. 5. It is obvious from Table 3 that the two molecules have almost the same VEE and OS values.

The close similarity between the calculated results for $n$-C$_4$H$_9$ONO and for (CH$_3$)$_2$C$_3$H$_5$ONO provides clues for elucidating our experimental observations described above. According to our *ab initio* calculations, the substitution of one of the two γ-H atoms in $n$-C$_4$H$_9$ONO with a methyl group yielding (CH$_3$)$_2$C$_3$H$_5$ONO does not substantially change the nature of the C−O−N=O core as this core remains in the *X-Y* plane with an almost unchanged electronic density distribution. Since the TR-FTIR emission spectra of the NO fragment as well as the UV absorption spectra for the two molecules are all closely related to the C−O−N=O core, it is not surprising that the obtained IR and UV spectra of $n$-C$_4$H$_9$ONO were found to bear a high degree of resemblance to those of (CH$_3$)$_2$C$_3$H$_5$ONO.

Considering that (1) the calculated VEE and OS values of the $S_1 \leftarrow S_0$ transition for the two molecules are quite close to one another, and (2) the temporal evolution of the IR emission of the nascent NO fragments from both molecules are nearly indistinguishable, we can anticipate that the 355-nm laser photolysis (and absorption) of the two species may undergo a similar photodissociation pathway. Figure 7 depicts the calculated potential energy curves as a function of the RO−NO bond length by freezing the optimized geometries of $n$-C$_4$H$_9$ONO and (CH$_3$)$_2$C$_3$H$_5$ONO in the ground $S_0$ state with only the RO−NO distance varied from 0.9 to 2.3 Å at an interval of 0.05 Å.[33] It is clearly seen in Fig. 7 that there is almost no appreciable difference for the $S_0$ (and $S_1$) potential energy curve between the two molecules, which in turn strongly suggests that the 355-nm photon (indicated by a vertical arrow in Fig. 7) used to bring both $n$-C$_4$H$_9$ONO and (CH$_3$)$_2$C$_3$H$_5$ONO from their $S_0$ state to $S_1$ state should guide the two molecules to an energetically very similar dissociation asymptote, thereby giving rise to NO ($X^2\Pi$) with a very similar rotational temperature and relative vibrational population.

Last but not least, we note that the absorption spectrum in the UV region of 300 – 450 nm for $(CH_3)_3CONO$ exhibits a rather large wavelength red-shift (~25 nm) with respect to those for $n$-$C_4H_9ONO$ and $(CH_3)_2C_3H_5ONO$,[7,30] as mentioned above. This can be readily understood by the fact that a drastic change of the electronic density distribution has recently been predicted to take place in the C−O−N=O core of $(CH_3)_3CONO$,[36] in contrast to the cases for $n$-$C_4H_9ONO$ and $(CH_3)_2C_3H_5ONO$ discussed in this paper.

## IV. CONCLUSION

We have investigated the 355-nm laser photolysis of $n$-butyl nitrite ($n$-$C_4H_9ONO$) and isoamyl nitrite (($CH_3)_2C_3H_5ONO$) using the TR-FTIR spectroscopic technique. We obtained the TR-FTIR emission spectra of the nascent NO fragment for the two molecules. Rovibrational simulations yielded almost identical results for the two molecules: an ~330 K rotational temperature and a relative vibrational population of ($v$ = 1) : ($v$ = 2) : ($v$ = 3 ) = (0.59) : (0.27) : (0.14), which indicates that the nascent NO photofragments were formed to be rotationally "cold" while vibrationally "hot" under our experimental conditions. We also found a close resemblance in the measured temporal profiles of the IR emission of NO and the recorded UV absorption spectra for the two molecules. Our experimental results implied that the substitution of one of the two γ-H atoms in $n$-$C_4H_9ONO$ with a methyl group yielding $(CH_3)_2C_3H_5ONO$ has a minor effect on the nature of the C−O−N=O core, which accounts for the subsequent dissociation dynamics. This is verified by our *ab initio* calculations using TD-DFT at the B3LYP/6-311G(d,p) level.

**ACKNOWLEDGEMENTS**

This work was supported by the National Natural Science Foundation of China (Grant Nos. 20673107 and 20873133), the National Key Basic Research Special Foundation of China (Grant No. 2007CB815203), the Chinese Academy of Sciences (KJCX2-YW-N24), and the Scientific Research Foundation for the Returned Overseas Chinese Scholars, Ministry of Education of China. The authors have enjoyed helpful discussions with Dr. J. Sun (Nanyang Technological University, Singapore) and the anonymous reviewer.

**Figure Captions**

**Fig. 1.**  (Color online) The partial TR-FTIR emission spectrum (1700 – 2000 cm$^{-1}$) of the nascent NO fragment recorded in the time interval of 1.0 – 2.0 $\mu$s after the 355-nm photolysis of $n$-C$_4$H$_9$ONO.  The upper trace is the experimental spectrum, while the lower one is the simulated spectrum.  The three vibrational bands are indicated by ticks.

**Fig. 2.**  (Color online) A comparison between the TR-FTIR emission spectra (1700 – 2000 cm$^{-1}$) of the NO fragments produced from $n$-C$_4$H$_9$ONO (upper trace) and (CH$_3$)$_2$C$_3$H$_5$ONO (lower trace, which is rescaled by 2× as its partial pressure was one half of that of $n$-C$_4$H$_9$ONO).

**Fig. 3.**  (Color online) Temporal profiles of the IR emission of the NO fragments produced from the 355-nm photolysis of $n$-C$_4$H$_9$ONO (triangles) and (CH$_3$)$_2$C$_3$H$_5$ONO (circles), which were obtained without the 300-kHz-bandwidth amplifier.

**Fig. 4.**  (Color online) Electronic absorption spectra of (a) $n$-C$_4$H$_9$ONO and (b) (CH$_3$)$_2$C$_3$H$_5$ONO in the UV region of 300 – 450 nm.  The dashed lines indicate the 355 nm that is the wavelength of the photolysis laser used in the experiment.

**Fig. 5.**  (Color online) The minimum energy geometries for $n$-C$_4$H$_9$ONO (left) and (CH$_3$)$_2$C$_3$H$_5$ONO (right).  Both structures were optimized at the B3LYP/6-311G(d,p) level.

**Fig. 6.**  (Color online) Profiles of the orbitals relating to the $S_1 \leftarrow S_0$ transition for $n$-C$_4$H$_9$ONO (left) and (CH$_3$)$_2$C$_3$H$_5$ONO (right).

**Fig. 7.**  (Color online) Potential energy curves of the $S_0$ and $S_1$ states for $n$-C$_4$H$_9$ONO and (CH$_3$)$_2$C$_3$H$_5$ONO, calculated at the B3LYP/6-311G(d, p) level, as a function of the RO−NO bond length.  The geometries were constrained to equilibrium positions of the ground $S_0$ state except for the RO−NO distance.

**Table Legends**

**Table 1**  Bond lengths and angles for B3LYP/6-311G(d,p) optimized geometries of $n$-C$_4$H$_9$ONO.

**Table 2**  Bond lengths and angles for B3LYP/6-311G(d,p) optimized geometries of (CH$_3$)$_2$C$_3$H$_5$ONO.

**Table 3**  Excitation orbitals (EO), vertical excitation energies (VEE), and oscillator strengths (OS) for the $S_1 \leftarrow S_0$ transition estimated by TD-DFT at the B3LYP/6-311G(d,p) level.

**Fig. 1.** (Ji *et al.*)

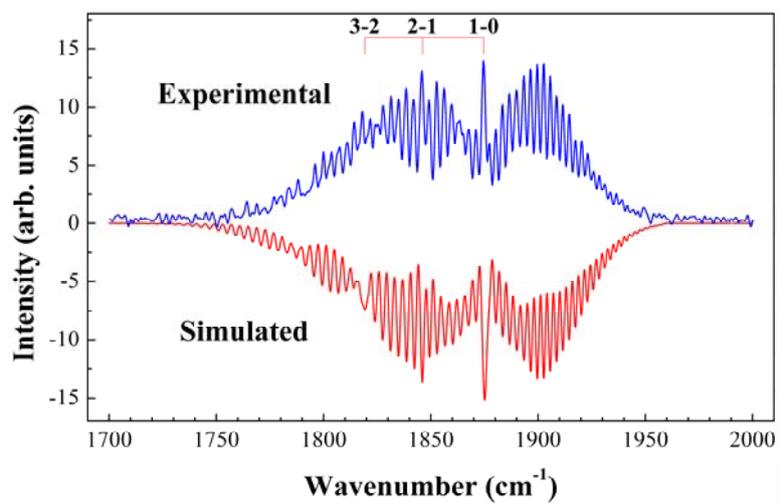

**Fig. 2.                    (Ji *et al.*)**

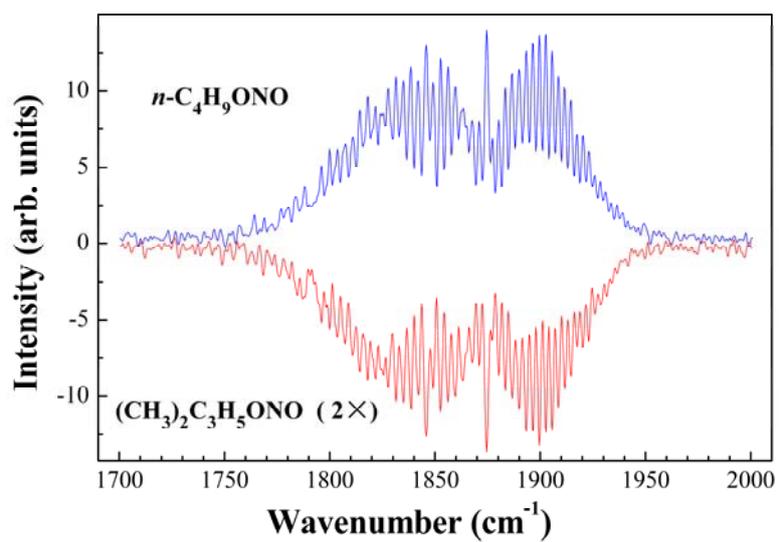

**Fig. 3.** (Ji *et al.*)

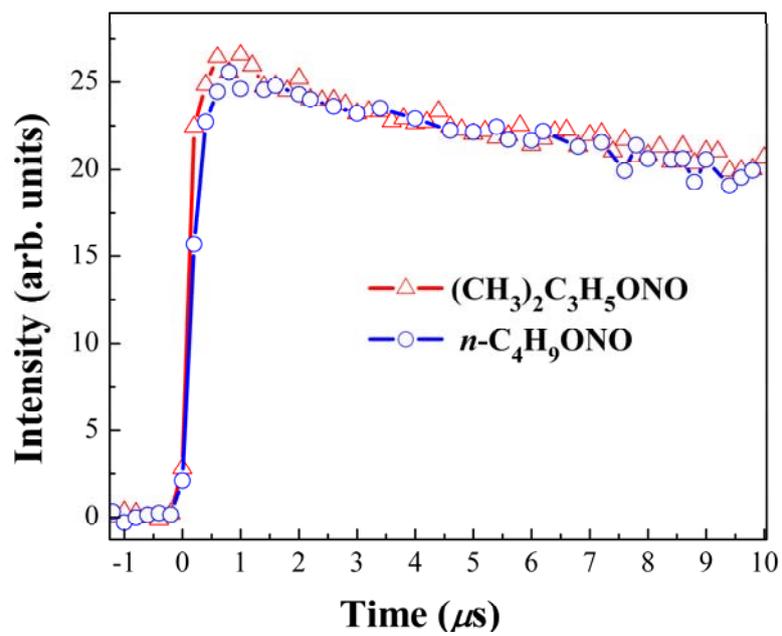

**Fig. 4.** (Ji *et al.*)

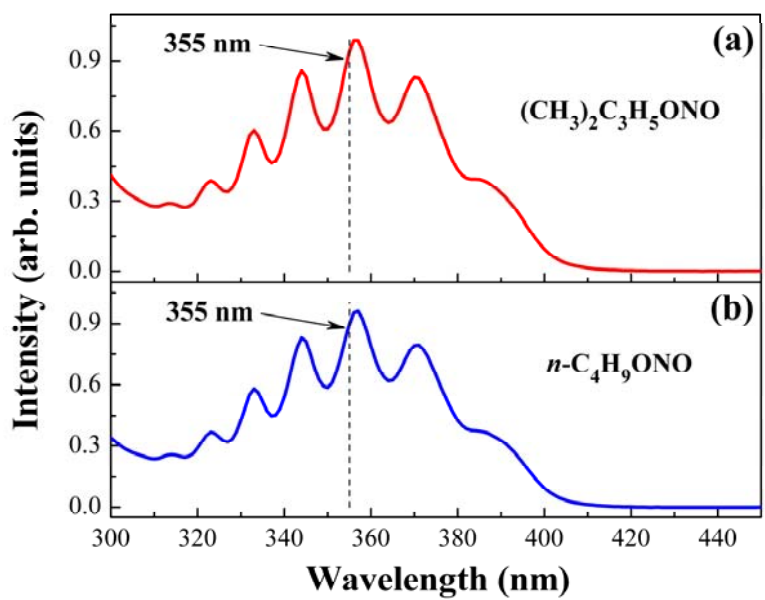

**Fig. 5.** (Ji *et al.*)

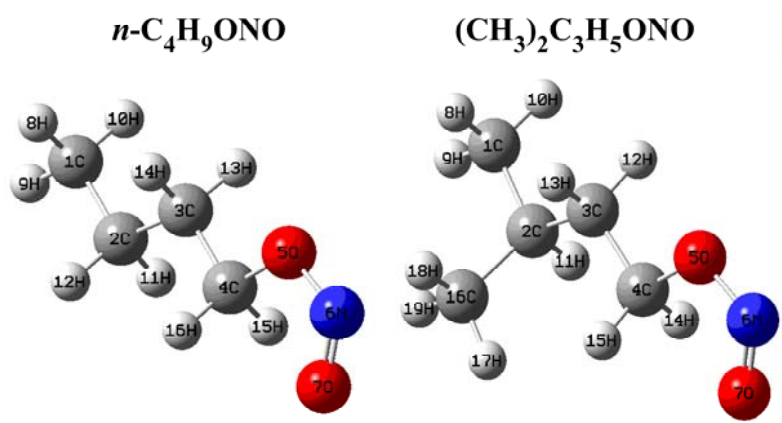

**Fig. 6.** (Ji *et al.*)

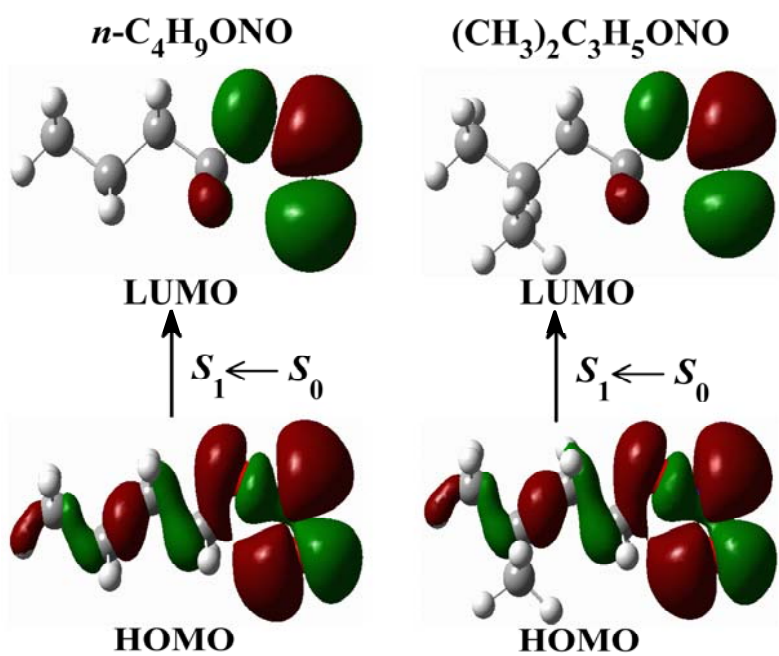

**Fig. 7.** (Ji *et al.*)

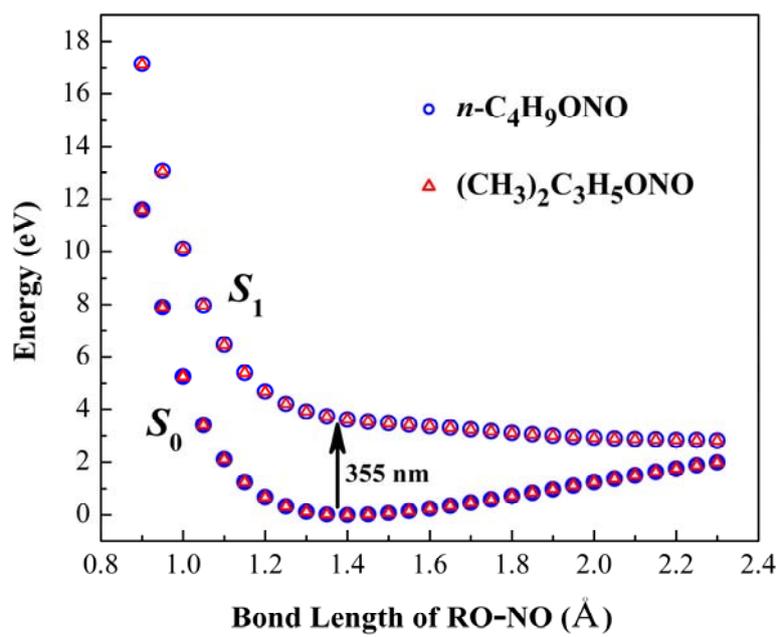

**Table 1**  Bond lengths and angles for B3LYP/6-311G(d,p) optimized geometries of *n*-$C_4H_9ONO$.

|  | Bond length (Å) |  | Bond angle (°) |
|---|---|---|---|
| $C_1$-$C_2$ | 1.531 | $O_7$-$N_6$-$O_5$ | 114.8 |
| $C_1$-$H_8$ | 1.094 | $N_6$-$O_5$-$C_4$ | 116.7 |
| $C_1$-$H_9$ | 1.093 | $O_5$-$C_4$-$C_3$ | 107.1 |
| $C_1$-$H_{10}$ | 1.094 | $O_5$-$C_4$-$H_{15}$ | 108.9 |
| $C_2$-$H_{11}$ | 1.096 | $O_5$-$C_4$-$H_{16}$ | 108.9 |
| $C_2$-$H_{12}$ | 1.096 | $H_{15}$-$C_4$-$H_{16}$ | 107.7 |
| $C_2$-$C_3$ | 1.534 | $C_4$-$C_3$-$C_2$ | 112.2 |
| $C_3$-$H_{13}$ | 1.095 | $O_7$-$N_6$-$O_5$-$C_4$ | 0.0 |
| $C_3$-$H_{14}$ | 1.095 | $N_6$-$O_5$-$C_4$-$C_3$ | 180.0 |
| $C_3$-$C_4$ | 1.519 | $O_5$-$C_4$-$C_3$-$C_2$ | 180.0 |
| $C_4$-$H_{15}$ | 1.095 | $N_6$-$O_5$-$C_4$-$H_{15}$ | 58.6 |
| $C_4$-$H_{16}$ | 1.095 | $N_6$-$O_5$-$C_4$-$H_{16}$ | -58.6 |
| $C_4$-$O_5$ | 1.448 | $C_4$-$C_3$-$C_2$-$C_1$ | 180.0 |
| $O_5$-$N_6$ | 1.400 | $H_{13}$-$C_3$-$C_4$-$H_{15}$ | 61.4 |
| $N_6$-$O_7$ | 1.181 | $H_{13}$-$C_3$-$C_4$-$H_{16}$ | -177.4 |

**Table 2** Bond lengths and angles for B3LYP/6-311G(d,p) optimized geometries of $(CH_3)_2C_3H_5ONO$.

|  | Bond length (Å) |  | Bond angle (º) |
|---|---|---|---|
| $C_1$-$C_2$ | 1.535 | $O_7$-$N_6$-$O_5$ | 114.8 |
| $C_1$-$H_8$ | 1.095 | $N_6$-$O_5$-$C_4$ | 116.6 |
| $C_1$-$H_9$ | 1.093 | $O_5$-$C_4$-$C_3$ | 106.6 |
| $C_1$-$H_{10}$ | 1.094 | $O_5$-$C_4$-$H_{14}$ | 108.9 |
| $C_2$-$H_{11}$ | 1.099 | $O_5$-$C_4$-$H_{15}$ | 108.6 |
| $C_2$-$C_{16}$ | 1.536 | $H_{14}$-$C_4$-$H_{15}$ | 107.8 |
| $C_2$-$C_3$ | 1.541 | $C_4$-$C_3$-$C_2$ | 113.9 |
| $C_3$-$H_{12}$ | 1.094 | $C_3$-$C_2$-$C_1$ | 110.0 |
| $C_3$-$H_{13}$ | 1.096 | $C_3$-$C_2$-$C_{16}$ | 112.6 |
| $C_3$-$C_4$ | 1.521 | $O_7$-$N_6$-$O_5$-$C_4$ | 0.0 |
| $C_4$-$H_{14}$ | 1.095 | $N_6$-$O_5$-$C_4$-$C_3$ | 179.7 |
| $C_4$-$H_{15}$ | 1.093 | $O_5$-$C_4$-$C_3$-$C_2$ | 175.8 |
| $C_4$-$O_5$ | 1.449 | $N_6$-$O_5$-$C_4$-$H_{14}$ | -59.4 |
| $O_5$-$N_6$ | 1.400 | $N_6$-$O_5$-$C_4$-$H_{15}$ | 57.8 |
| $N_6$-$O_7$ | 1.182 | $C_4$-$C_3$-$C_2$-$C_1$ | -172.4 |
| $C_{16}$-$H_{17}$ | 1.093 | $C_4$-$C_3$-$C_2$-$C_{16}$ | 63.6 |
| $C_{16}$-$H_{18}$ | 1.095 | $H_{13}$-$C_3$-$C_4$-$H_{15}$ | 57.8 |
| $C_{16}$-$H_{19}$ | 1.093 | $H_{13}$-$C_3$-$C_4$-$H_{14}$ | 179.6 |

**Table 3**  Excitation orbital (EO), vertical excitation energies (VEE), and oscillator strengths (OS) for the $S_1 \leftarrow S_0$ transition estimated by TD-DFT at the B3LYP/6-311G(d,p) level.

| Molecule | Transition | EO | VEE (eV) | OS |
|---|---|---|---|---|
| $n$-C$_4$H$_9$ONO[a] | $S_1 \leftarrow S_0$ | HOMO→LUMO | 3.6059 | 0.0021 |
| (CH$_3$)$_2$C$_3$H$_5$ONO | $S_1 \leftarrow S_0$ | HOMO→LUMO | 3.6025 | 0.0020 |

[a]Note: The results given in this row for $n$-C$_4$H$_9$ONO are in good agreement with those reported in Ref. [2].